\documentclass[conference]{IEEEtran}
\IEEEoverridecommandlockouts
\usepackage{cite}
\usepackage{amsmath,amssymb,amsfonts}
\usepackage{graphicx}
\usepackage{textcomp}
\usepackage{xcolor}
\usepackage{authblk}
\usepackage{algorithm,algpseudocode}
\usepackage{pgfplots}
\def\BibTeX{{\rm B\kern-.05em{\sc i\kern-.025em b}\kern-.08em
    T\kern-.1667em\lower.7ex\hbox{E}\kern-.125emX}}
    \pgfplotsset{width=7cm,compat=1.8}
\begin{document}

\title{An Analysis of Fog Computing Data Placement Algorithms}


\author[1]{Daniel Maniglia Amancio da Silva }
\author[1]{Godwin Asamooning}
\author[1]{Hector Orrillo}
\author[2]{Rute C. Sofia}
\author[3]{Paulo M. Mendes} 

\affil[1]{COPELABS - University Lusofona, \footnote{ Campo Grande 388, 1700-097 Lisboa, Portugal} (e-mail: daniel.maniglia@hotmail.com)}
\affil[2]{fortiss \footnote{fortiss -  research institute of the Free State of Bavaria for software intensive systems and services. Guerickerstr. 25, 80634 München} (e-mail: sofia@fortiss.org)}
\affil[3]{Airbus Central Research and Technology, Germany (e-mail: paulo.mendes@airbus.com)}

\maketitle


\begin{abstract}
This work evaluates three Fog Computing data placement algorithms via experiments carried out with the iFogSim simulator. The paper describes the three algorithms (Cloud-only, Mapping, Edge-ward) in the context of an Internet of Things scenario, which has been based on an e-Health system with variations in applications and network topology. Results achieved show that edge placement strategies are beneficial to assist cloud computing in lowering latency and cloud energy expenditure.
\end{abstract}

\section{Introduction}
\emph{Cloud computing} flexibly organises distribution of network resources and data storage as well as processing \cite{hu2017secure}. \emph{Fog Computing}, also known as \emph{Edge Computing} \cite{shi2016edge}, is a set of paradigms that assist computation, networking, and storage between the edges of the network and the Cloud. 
The main goal of Fog Computing is to extend the Cloud capabilities to the edges of the network, thereby supporting real-time data processing and latency sensitive applications. In Fog Computing, resources are dynamically distributed across the Cloud and network elements based on \emph{Quality of Service (QoS)} requirements \cite{sarddar2018refinement}. Fog computing can assist in lowering latency, by allowing data and computation of data to be placed closer to the end-user. The idea in this context is that Fog devices such as gateways, switches, routers, can store/serve application modules before they are sent to the Cloud.
As computing requirements of several Internet services, such as \emph{ Internet of Things (IoT)}, Augmented Reality, video streaming adaptation, keep on increasing, there are several challenges faced by Cloud computing. 
For instance, end-to-end QoS of video streaming in mobile networks has been improved by introducing in-network computation (QoS mapping and adaptation) at the network edges, aiming to map multimedia flows to the most suitable network service class across heterogeneous mobile networks, and to adapt the session to the current network conditions \cite{cerqueira2008qos} \cite{cerqueira2007qos}.
However, management of processing and storage at the different hierarchical levels together with the varied user application classes and requirements, has become increasingly complex and calls for the use of data placement algorithms to assist in evaluating such distributed processing and storage capacities as well as application requirements for efficient data placement \cite{bittencourt2017mobility}. 

This work focuses on the evaluation of three data placement algorithms, namely, Cloud-only, Edge-ward, Mapping. The algorithms are described and evaluated via simulations carried out with the iFogSim simulator. The performance evaluation concerns execution time and energy consumption from a Cloud perspective.

The paper is structured as follows. Section \ref{Related} describes related work and the contributions of this work towards related literature. Section \ref{Problem} provides background on Fog Computing. Section \ref{Proposed} explains the selected data placement algorithms, while section \ref{results} evaluates and validates the selected algorithms based on iFogSim. The paper concludes with section \ref{conclusions}, where directions for future work are also debated.

\section{Related Work} 
\label{Related}
Prior research has shown that the selection of specific entities in operator networks, which are commonly used in large-scale distributed environments, has a significant impact on relevant performance metrics, such as network utilisation, or end-to-end latency \cite{rizou2010solving}\cite{schilling2011efficient}. Therefore, with a bright future at hand to provide satisfactory computing performance due to the massive growth in IoT devices, together with boundless prospective IoT applications under next-generation aiming to contribute positively to real time processing of data at locations closer to the end user. There is an urgent need for collaboration among IoT devices, as they are constrained and the respective services require intensive computing capacity and persistent data processing abilities. IoT cloud-based services may not be able to propel the prospective applications envisioned \cite{pham2019joint}. 

So, as Fog nodes are distributed across the network edge, they can cooperate to support distributed processing and storage, in cooperation with the Cloud. However, Fog currently requires some type of resource broker to manage computing resources and also to effectively schedule computational functions across its hierarchical structure \cite{pham2016towards}. Thus, scheduling of computational functions in Fog Computing requires mechanisms that combine features from Cloud computing with opportunistic networking. For instance, Olaniyan et al. describe the model for opportunistic edge computing, where available resources are more volatile, given that they are also based on end-user controlled devices \cite{olaniyan2018opportunistic}.

Several related work analysed scheduling and task allocation performance aspects in Fog computing. Singh et al. propose a combination of virtual machine allocation and virtual machine selection to optimise the scheduling of tasks assigned to Cloud processes \cite{singh2019dynamic}. 
Rahbari et al. review scheduling strategies and parameters using the greedy knapsack-based scheduling (GKS) algorithm to seek efficient allocation of resource modules in fog networks \cite{rahbari2019low}.  

The combination of the optimal placement of data blocks and task scheduling results in improved response time \cite{li2019joint} as proposed by Li et al. Their method concerns the "popularity" of data blocks, the storage capacity of different devices, and server replacement ratios. In a subsequent work, Li et al. propose a parallel virtual queuing algorithm for buffering different tasks in Fog nodes \cite{li2019resource}. The algorithm combines an adaptive queuing weight resource allocation policy with a task buffering and offloading policy. 

A resource-aware algorithm for placement of data analytics platform in Fog computing is also proposed \cite{taneja2016resource}. The platform is assumed to execute services on a virtual machine and expect to actively deploy the analytical platform to trigger where to optimally run a service on the Fog or the Cloud, thus reducing network costs and response time to user requests.  
From an operational perspective, there are several works focusing on initial evaluations of Fog Computing models. For instance, Bittencourt et al. evaluate scheduling for Fog computing specifically focusing on the challenge of mobility \cite{bittencourt2017mobility}.  Gupta et al. provide iFogSim, a simulator intended to assist resource management evaluation in Internet of Things (IoT) scenarios \cite{gupta2017ifogsim}. In addition to describing the simulator their work exemplifies the simulation of different placement strategies on network parameters of delay, energy consumption and cost.

Our work considers these same placement strategies, providing a first evaluation for a specific case of more heterogeneous IoT environments derived from personal IoT, and having as performance parameters latency and energy consumption in the cloud.

\section{Background} \label{Problem}
\subsection{Fog Computing Architecture}
\begin{figure}[ht]
    \centering
    \includegraphics[width=0.4\textwidth]{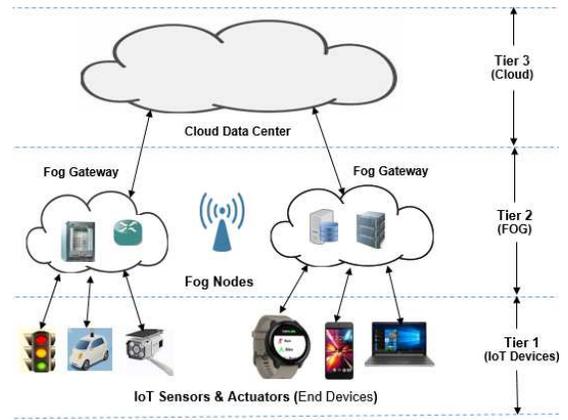}
    \caption{Fog-Cloud computing architecture.}
    \label{ARQUITETURA}
\end{figure}

A Fog computing architecture is intended to improve both network management and storage and application processing \cite{iorga2018fog}\cite{choudhari2018prioritized}\cite{maiti2018qos}. For that purpose, Fog architectures integrate mechanisms that can better distribute resources across a specific infrastructure.  Figure \ref{ARQUITETURA} illustrates such a networking architecture, where Layers represent Tier levels. 

\textbf{Tier 1} integrates IoT \emph{field-level} devices, such as sensors and actuators. These are data sources,  devices that capture and distribute data to other Tier devices, same Tier, or next Tier level. 

\textbf{Tier 2} (FOG) integrates IoT devices coined as \emph{Fog nodes} \cite{tordera2016fog}. IoT hubs and gateways that gather data and process information fall into this category. Gateways are responsible for translating communication protocols and assist in data transmission to different network segments. The Tier 2 level includes also devices such as routers and \emph{Access Points (AP)}. Fog nodes are arranged in a hierarchical way and communication is only possible between a parent-child pair in the hierarchy. Given that these devices are in the edges of the network, often located in Customer Premises, Fog nodes often have limited resources.

\textbf{Tier 3} (CLOUD) devices often have a significantly higher amount of resources. These are, for instance, virtual machines in data centers.

Data processing occurs in all of the three Tiers thus consuming network and computational resources such as energy, memory, CPU, network capacity. 
Data Placement algorithms are responsible for supporting the distribution of services, data, and applications to specific devices, and Fog layers. Different algorithms result in different execution placement and allocation of network resources.

\section{Selected Data Placement Algorithms} \label{Proposed}
To evaluate the performance behavior of different data placement strategies, this work has considered three different algorithms.
The selected Data Placement Algorithms are: Edge-ward, Cloud-only and Mapping.

\subsection{Edge-ward Algorithm}
Edge-ward  \cite{gupta2017ifogsim} is based on a \emph{First Come-First Served (FCFS)} strategy, and therefore results in placing data as close as possible to the edge of the network, on Fog nodes. This algorithm has been selected as it represents a placement strategy focused on the Edge only.

If a specific Fog node cannot serve the requirements of an application, Edge-ward selects additional Fog devices. 
The algorithm  creates tuples of devices representing the paths via which \emph{"application modules"} (services) are executed.

Application requests are answered based on the order in which they arrive until there are available computing resources at each hierarchical level. If the Fog device selected to run an application does not have available computational capacity, then the algorithm searches for a Fog device with capacity at the top layer of the network topology hierarchy.

In other words, application requests are answered based on the order in which they arrive until there are available computing resources at each hierarchical level. The algorithm was chosen because it presents better performance when planning resources (specifically CPU) of fog devices in a hierarchical way. If a fog device is unable to meet the requirements of an application module, then it can be scheduled in the Cloud. Algorithm 1 provides the pseudo-code for Edge-ward, where:

\textit{p : path}

\textit{d : device}

\textit{w : modules}

\textit{$\theta$:  selected$\quad$ module}



\begin{algorithm}
\caption{-- Edge-ward} 
\label{Edge-ward}
\begin{algorithmic}[1] 
\While{$p\quad\epsilon\quad$ PATHS}\Comment{Across all paths}
\State \emph{placeModules} $:$= \{\} \Comment{device list}

\While{Fog device $d\quad\epsilon\quad $p}
    \{\}\Comment{leaf-to-root traversal}
     \State \emph{modulesToPlace} $:$= \{\}
        \While{module \emph{w} $\quad \epsilon \quad $app} 
        \Comment{find modules \quad realy for placement on device d}
            \If{all predec. of \emph{w} are in placedModules}\Comment{if all predecessors are placed}      \State add \emph{w} to modulesToPlace 
                \EndIf
        \EndWhile
        \While{$module \quad \theta \quad \epsilon \quad modulesToPlace$} 
\If{$d \quad has \quad instance \quad of \quad \theta \quad as~\theta'$}
\If{$CPU \theta >= CPU$d}
\Comment{device d does not have CPU capacity to host $\theta$}
    \State $\theta":= merge(\theta,\theta') $
    \State $f := parent(d) $
        \While{$CPU \theta" >= CPUf$}
        \Comment{find device north of d for hosting $\theta$}
            \State $f := parent(f) $
        \EndWhile
    \State$Place \quad \theta" \quad on \quad device \quad f$
    \Comment{device $\quad f \quad can \quad host \theta"$}
        \State$add \quad \theta \quad to \quad placedModules$
    \ElsIf{
    \State$Place \quad \theta \quad on \quad device \quad d$  \Comment{device $\quad d \quad can \quad host \theta$}
    \State add $\quad \theta \quad to \quad placedModules$}
    \EndIf
    \ElsIf{no device north (near the cloud) of \textit{d} has an instance $\theta$}
    
    \If{$CPU \theta <= CPUd$}
        \Comment{if not, will be handled by subsequent iterations}
        \State$Place \quad \theta \quad on \quad device \quad d$
       
        \State$add \quad \theta \quad to \quad placedModules$
\EndIf
\EndIf
        \EndWhile
    \EndWhile
\EndWhile
\end{algorithmic}
\end{algorithm}

\subsection{Cloud-only Algorithm}

This algorithm is based on the traditional implementation executed in the Cloud and uses a \emph{delay priority strategy} \cite{gupta2017ifogsim}.
Cloud-only placement assumes that all application modules run in data centers. Therefore, this algorithm follows the traditional situation where sensors capture data; such data is processed on the cloud; the cloud sends information to actuators if required.
Cloud-only placement is provided in Algorithm 2. 

\begin{algorithm}
 \caption{-- Cloud-only}
    \label{onlycloud}
    \begin{algorithmic}[1] 

\While{$p\quad\epsilon\quad$ PATHS} \Comment{Across all paths}
\State \emph{placeList} $:$= \{\} \Comment{device list}
    \While{Fog device $d\quad\epsilon\quad $p} \Comment{way}
        \While{$module \quad \emph{w} \quad \epsilon \quad $app} 
                \If{all predec. of \emph{w} are placed}
                \State add \emph{w} to placeList 
                \EndIf
                \EndWhile
                \While{$module \quad \theta \quad \epsilon \quad placeList$} 
                    \State $Place\quad \theta\quad on \quad device \quad CLOUD$
                \EndWhile
            \EndWhile
\EndWhile
    \end{algorithmic}
\end{algorithm}

\subsection{Mapping Algorithm}
The Mapping algorithm relies on a \emph{concurrent} strategy. Application requests are mapped preferably to Fog devices, independently of their application capabilities and requirements \cite{bittencourt2017mobility}.
Therefore, if the CPU capacity of the selected Fog device does not suffice to serve an application requirements, then Mapping forms a processing queue on the Fog node.

The algorithm receives as a parameter the lists of possible paths via which the application can be executed following a leaf-to-root traversal.
The Mapping pseudo-code is provided in Algorithm 3.

\begin{algorithm}
\caption{-- Mapping} 
\label{Mapping}
\begin{algorithmic}[1] 
\While{$p\quad\epsilon\quad$ PATHS} \Comment{Across all paths}

\State \emph{placeList} $:$= \{\} \Comment{device list}
    \While{Fog device $d\quad\epsilon\quad $p} \Comment{way}
        \While{module \emph{w} $\quad \epsilon \quad $app} 
                \If{all predec. of \emph{w} are placed}
                \State add \emph{w} to placeList 
                \EndIf
        \EndWhile
   \While{$module \quad \theta \quad \epsilon \quad placeList$} 

\If{$ \theta \quad ahead \quad place\quad on \quad d \quad \epsilon \quad p$}
    \State $d := Device $ 
    \State$Place \quad \theta \quad on \quad device \quad d$
    \EndIf
        \EndWhile
    \EndWhile
\EndWhile
\end{algorithmic}
\end{algorithm}

\section{Performance Evaluation} \label{results}
\subsection{IFogSim IoT Simulator} 
The algorithms have been evaluated via simulations carried out with the iFogSim Simulator \cite{gupta2017ifogsim}. 
iFogSim enables the simulation of resource management and application scheduling policies. 
The simulator integrates different modules. 
\emph{IoT devices} model field-level devices, and are defined based on specific parameteres, such as CPU usage.

\emph{IoT Applications} build the data processing elements based on specific \emph{Application Models}, which are objects that simulate tupples between sources and destinations. For the specific case of the selected scenario, each application corresponds to a Fog tupple, i.e., a session established to transmit data between sensors and cloud, or sensors and Fog Devices (e.g., IoT gateway, smartphone). 
Each tupple is characterized by the following attributes: i) CPU length, in millions of instructions per second (MIPS); ii) \emph{Network Length (NW Length)}, corresponding to the bandwidth occupied by the active application session; iii) average inter-arrival time.

\emph{Fog devices} model Fog nodes, i.e., devices capable of hosting application modules. 
\emph{Resource management} is a core component of IFogSim and is composed by \emph{Placement} and \emph{Scheduler}.
This modular structure allows the deployment of different scenarios, as shall be explained in the next sections. 

\subsection{Underlying Experimental Scenario}

\begin{figure}[ht]
    \centering
    \includegraphics[width=0.3\textwidth]{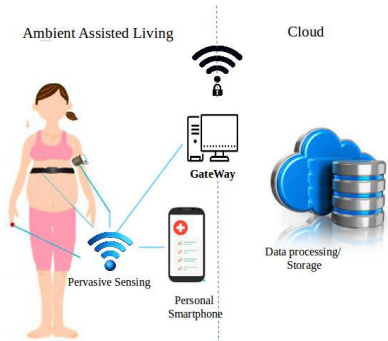}
    \caption{IoT Smart Health Scenario.}
    \label{Scenario}
\end{figure}

To evaluate the different algorithms, this paper considers a scenario where a user relies on a personal IoT Smart Health kit. A representation of such a scenario is provided in Figure \ref{Scenario}.

Such a kit is composed of a smartphone with a specific application and multiple sensors installed in the Customer Premises. The end-user carries a smartphone with multiple sensors (e.g., accelerometer, GPS, Wi-Fi, heart rate). Additional sensors collecting data concern, for instance, temperature, power usage at home. Co-located with an AP is an IoT gateway, which provides support for communication to and from the cloud. Some of the data collected is regularly sent to the cloud, to a server on the clinic where the patient is being followed, with the purpose of raising awareness to possible health issues that may arise, e.g., a high heart rate.

The different sensors provide real-time data measurement both to the gateway and smartphone healthcare kit application. Such real-time data may be locally stored and filtered; depending on the type of data, whereabouts of the user \cite{sofia2015tool}, as well as other conditions (e.g., threshold alerts), the data can be kept on the smartphone, IoT gateway, or sent to the cloud for further processing. For instance, temperature sensors may rely upon a communication protocol such as the \emph{Advanced Message Queuing Protocol (AMQP)} \cite{amqp} or the \emph{Message Queuing Telemetry Transport (MQTT)} \cite{mqtt} to send temperature updates periodically to the IoT gateway, where a broker resides. Such information is sent to the application on the smartphone (subscriber) and to the server-side application on the clinic (subscriber), as well as to relatives of the patient that have also the same smartphone application (subscribers). On the smartphone, the application captures data over multiple sensors to evaluate health aspects but also to raise awareness to other aspects that may indicate issues, e.g., isolation of the user, lack of movement \cite{NSense}.

\subsection{Experimental Settings}
 The purpose of the evaluation is to compare the performance of the three selected data placement algorithms in terms of execution time, and energy consumption on the Cloud. 
To evaluate the algorithms, the iFogSim scenario selected is based on parameters that have been thought to fulfil the requirements of the scenario described.

\begin{figure}[ht]
    \centering
    \includegraphics[width=0.3\textwidth]{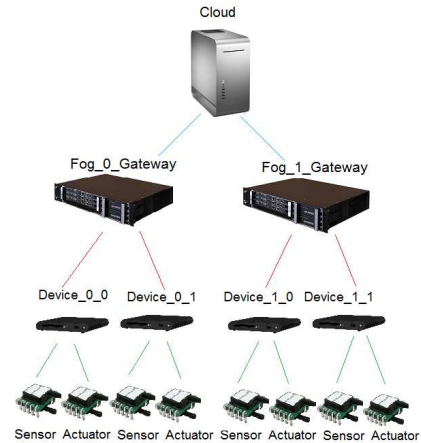}
    \caption{Topology 1.}
    \label{top1}
\end{figure}

\begin{figure}[ht]
    \centering
    \includegraphics[width=0.4\textwidth]{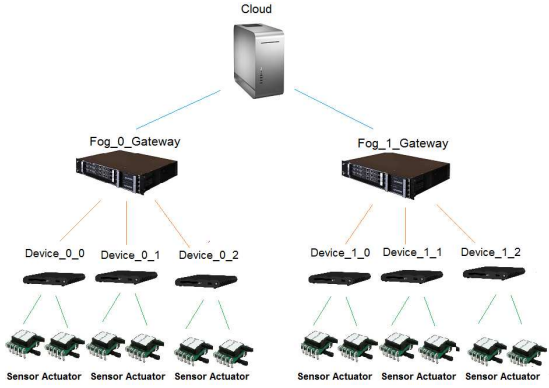}
    \caption{Topology 2.}
    \label{top2}
\end{figure}

\begin{figure}[ht]
    \centering
    \includegraphics[width=0.5\textwidth]{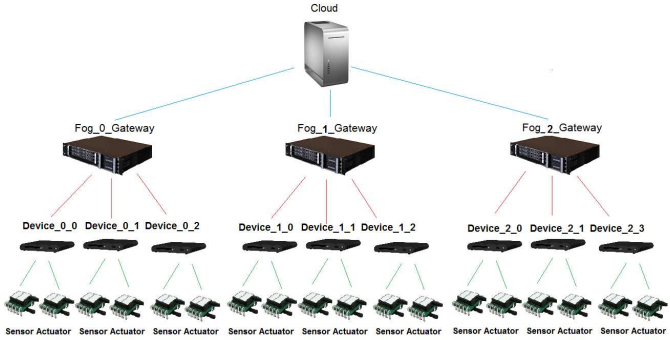}
    \caption{Topology 3.}
    \label{top3}
\end{figure}

The scenario integrates three different Fog topologies respectively represented by Figures \ref{top1}, \ref{top2}, \ref{top3}. 
Topology (\emph{Top1}), illustrated in  Figure \ref{top1},  consists of one cloud, two gateways, two Fog devices for each gateway, and one actuator and one sensor for each Fog device.

The second topology (\emph{Top2}), illustrated in Figure \ref{top2}, consists of one cloud-based server, three gateways, two Fog devices for each gateway, and one actuator and one sensor for each Fog device. 

The third topology (\emph{Top3}), illustrated in Figure \ref{top3}, consists of one cloud, two gateways, three Fog devices for each gateway, and one actuator and one sensor for each Fog device.

The Fog devices attributes are described in Table \ref{FogDevices}. There are three different types of devices in the proposed scenario: Cloud, Gateway, and SmartPhone. Each device is characterized by the capacity of CPU processing, RAM, upload and download bandwidth (UpBW and DownBW in KBytes), hierarchical level in the network and also rate per MIPS (cost rate per MIPS used). The selected values have been based on related literature \cite{gupta2017ifogsim} \cite{mahmudmodelling} \cite{rahman2018performance}. 

The gateway device simulates the use of the "intel core 2" processor and the smartphone simulates the use of a "cortex-A53" processor and the RAM, CPU (GHz) and storage capacity values refer to these processor models. 
The processing capacity in MIPS is an approximate ratio to the value in GHz. \footnote{http://www.roylongbottom.org.uk/cpuspeed.htm.} 


\begin{table}[]
\caption{Fog devices attributes.\label{FogDevices}}
\centering
\begin{tabular}{|c|c|c|c|c|}
\hline
 & \textbf{Cloud} & \textbf{Gateway} & \textbf{SmartPhone} &
 \textbf{Sensor} \\ \hline
\begin{tabular}[c]{@{}c@{}}CPU \\ (MIPS)\end{tabular} & 44800 & \begin{tabular}[c]{@{}c@{}}2600 - \\ 7000\end{tabular} & \begin{tabular}[c]{@{}c@{}}1000-\\ 2800  \end{tabular} & 0  \\ \hline
\begin{tabular}[c]{@{}c@{}}CPU\\  (GHz)\end{tabular} & 130 & 2.4 & 1.6 & 0  \\ \hline
\begin{tabular}[c]{@{}c@{}}RAM\\  (GB)\end{tabular} & 40 & 4 & 1 & 0 \\ \hline
UpBW & 100 & 10000 & 10000 & 1000\\ \hline
DownBW & 10000 & 10000 & 10000 & 1000\\ \hline
Level & 0 & 1 & 2 & 3\\ \hline
\begin{tabular}[c]{@{}c@{}}Rate per\\  MIPS\end{tabular} & 0,01 & 0 & 0 & 0\\ \hline
\begin{tabular}[c]{@{}c@{}}Storage\\  (GB)\end{tabular} & 12500 & 500 & 32 & 0 \\ \hline
\end{tabular}
\end{table}

Moreover, the simulator allows also for the modelling of transmission delay between the different devices, as provided in Table \ref{delay}. 
\begin{table}[]
\caption{Network link latency characterization.}
\label{delay}
\centering
\begin{tabular}{|c|c|c|}
\hline
\textbf{Source} & \textbf{Destination} & \textbf{\begin{tabular}[c]{@{}c@{}}Link delay (ms)\\ (milliseconds)\end{tabular}} \\ \hline
Sensor & Smartphone & 2 \\ \hline
Sensor & Gateway & 5 \\ \hline
Smartphone & Gateway & 20 \\ \hline
Smartphone & Cloud & 50 \\ \hline
Gateway & Cloud &  100 \\ \hline
\end{tabular}
\end{table} 

The simulation consists of extracting "runtime" (execution time) results, cloud CPU consumption and execution placement identification (Cloud, Gateway or Fog Device) in a number of runs. Each simulated scenario results from a combination of the three different scenarios with three different applications described next.

\subsection{Application Characterization} \label{application}

\begin{table}[ht]
\caption{Application characterization.\label{ApplicationsTable}}
\centering
\begin{tabular}{|c|c|c|c|}
\hline
\textbf{\small Application} & \textbf{\begin{tabular}[c]{@{}c@{}} CPU \\ (MIPS) \end{tabular}} &  \textbf{\begin{tabular}[c]{@{}c@{}} NW \\ (KBytes)  \end{tabular}}
&  \textbf{\small IA (ms)} \\
\hline
\small 1 & \small  1000 & \small  1 & \small 1000 \\ \hline
\small 2 & \small 5000 & \small 1000 & \small 50 \\ \hline
\small 3 & \small 10000 & \small 7000 & \small 20 \\ \hline
\end{tabular}
\label{CPUConsuptions}
\end{table}

The simulations developed count with three different iFogSim application models. Table \ref{CPUConsuptions} provides the attributes for each application, where CPU stands for CPU usage in MIPS; NW corresponds to the network capacity used by the application, in KBytes, and IA corresponds to inter-arrival time of the application tupples in milliseconds.

Application 1 corresponds to a low intensity application model. This can be, for instance, data periodically sent by environmental sensors to devices around (Fog and Cloud). It has been characterised as requiring 1000 MIPS of CPU, and a NW Length of 1 Kbyte. Inter-arrival time is long, having been set to 1 second (s).

Application 2 models a medium-intensity tupple, e.g., data being periodically sent from an app on a smartphone to devices around or to the Cloud. It has been characterised as consuming 5000 MIPS, a NW length of 1 MByte, and an average inter-arrival time of 50 ms.

Application 3 models a high intensity tupple, e.g., data being periodically sent from the application to a medical clinic. It has therefore been characterised as consuming 10000 MIPS and occupying a NW length of 7 MBytes. The inter-arrival time has been set to 20 ms.

\subsection{Assumptions}
The following assumptions have been considered in the simulations:

\begin{enumerate}
\item The number of applications are directly linked with the number of available sensors. One application always starts at a sensor. The data placement algorithm selects where to perform the computation of such application and provides results to an actuator.  
\item The StorageModule is always executed in the Cloud. The ClientModule is placed on devices according to the data placement algorithm used. 
\item The "Execution Time" is the sum of the execution times of all modules of the executed applications. This includes the processing time of the placement algorithm and the latency between the devices used, from the output of the sensor to the return of the actuators.  
\end{enumerate}

\subsection{Results}

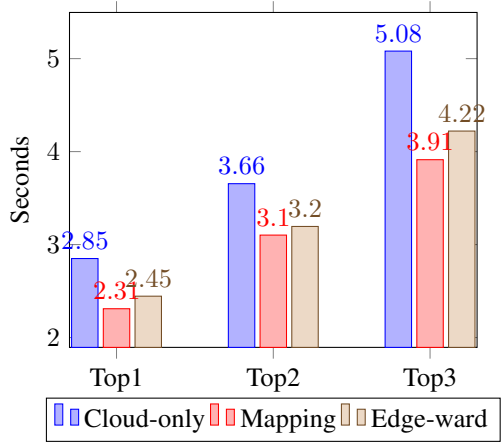
\begin{figure}[ht]
\centering
\begin{tikzpicture}
\begin{axis}[
    ybar,
    enlargelimits=0.15,
    legend style={at={(0.5,-0.15)},
      anchor=north,legend columns=-1},
    ylabel={Seconds},
    symbolic x coords={Top1,Top2,Top3},
    xtick=data,
    nodes near coords,
    nodes near coords align={vertical},
    ]
\addplot coordinates {(Top1,2.850) (Top2,3.655) (Top3,5.081)};
\addplot coordinates {(Top1,2.310) (Top2,3.102) (Top3,3.913)};
\addplot coordinates {(Top1,2.445) (Top2,3.195) (Top3,4.221)};
\legend{Cloud-only,Mapping,Edge-ward}
\end{axis}
\end{tikzpicture}
\caption{Application 1 - Execution Time.} \label{chart1}
\end{figure}

\begin{figure}[ht]
\centering

\begin{tikzpicture}
\begin{axis}[
    ybar,
    enlargelimits=0.15,
    legend style={at={(0.5,-0.15)},
      anchor=north,legend columns=-1},
    ylabel={Seconds},
    symbolic x coords={Top1,Top2,Top3},
    xtick=data,
    nodes near coords,
    nodes near coords align={vertical},
    ]
\addplot coordinates {(Top1,2.960) (Top2,3.745) (Top3,5.103)};
\addplot coordinates {(Top1,2.513) (Top2,3.502) (Top3, 4.388)};
\addplot coordinates {(Top1,2.695) (Top2,3.240) (Top3,4.543)};
\legend{Cloud-only,Mapping,Edge-ward}
\end{axis}

\end{tikzpicture}
\caption{Application 2 - Execution Time.} \label{chart2}
\end{figure}
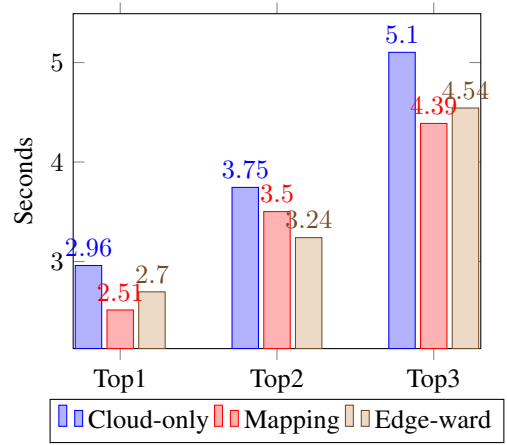

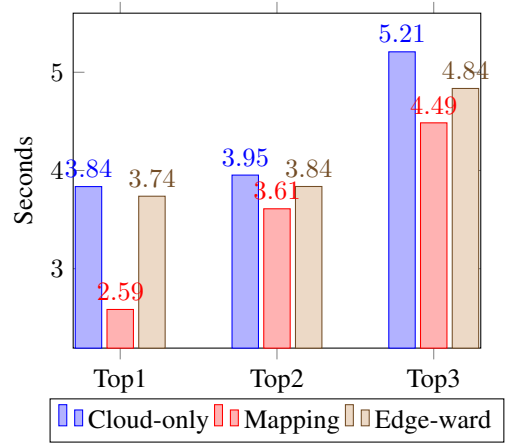
\begin{figure}[ht]
\centering
\begin{tikzpicture}
\begin{axis}[
    ybar,
    enlargelimits=0.15,
    legend style={at={(0.5,-0.15)},
      anchor=north,legend columns=-1},
    ylabel={Seconds},
    symbolic x coords={Top1,Top2,Top3},
    xtick=data,
    nodes near coords,
    nodes near coords align={vertical},
    ]
\addplot coordinates {(Top1,3.836) (Top2,3.953) (Top3,5.208)};
\addplot coordinates {(Top1,2.585) (Top2,3.610) (Top3,4.485)};
\addplot coordinates {(Top1,3.738) (Top2,3.837) (Top3,4.835)};
\legend{Cloud-only,Mapping,Edge-ward}
\end{axis}

\end{tikzpicture}
\caption{Application 3 - Execution Time.} \label{chart3}
\end{figure}

Figures \ref{chart1}, \ref{chart2}, \ref{chart3} provide execution time results for the three applications.

In terms of application intensity, the three algorithms exhibit similar performance. Execution time increases for Application 3 as expected, given that this is an example of a highly intensive application.
\subsubsection{Execution Time}
Comparing the execution time of the different algorithms, the Cloud-only algorithm always results in a higher execution time. These results can be is justified by the execution in the cloud, where it runs an execution path with greater latency (see Table \ref{delay}).

For low-intensity and average intensity applications, Edge-ward performs slightly worse than the Mapping algorithm. The execution time difference becomes significantly higher for high-intensity applications. 

Edge-ward is more sensitive to changes in topology, given that the higher number of Fog devices results in a higher number of application executions. Once the application needs exceed the capacity of Fog Devices, Edge-ward runs the application modules in CLOUD devices, thus increasing latency when compared to the Mapping algorithm.

Overall, the Mapping algorithm results in the lowest execution times. This is due to the Mapping algorithm forming processing queues if Fog devices do not have enough resources to accomodate applications. 

Independently of the selected application and of the selected algorithms, a common aspect is that an increase in Fog devices increases the amount of execution of the application modules thus increasing overall latency.

\subsubsection{Cloud Energy Consumption }
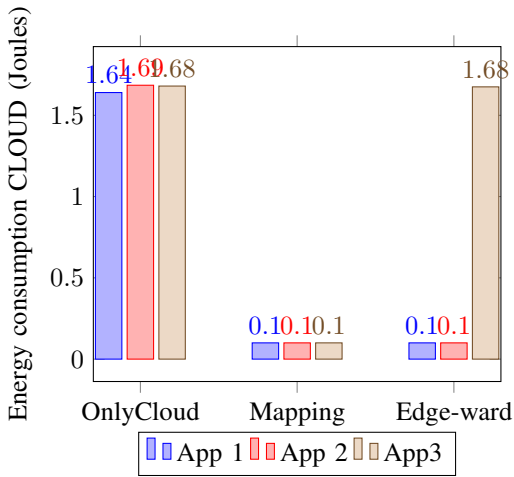
\begin{figure}[ht]
\centering
\begin{tikzpicture}
\begin{axis}[
    ybar,
    enlargelimits=0.15,
    legend style={at={(0.5,-0.15)},
      anchor=north,legend columns=-1},
    ylabel={Energy consumption CLOUD (Joules)},
    symbolic x coords={OnlyCloud,Mapping,Edge-ward},
    xtick=data,
    nodes near coords,
    nodes near coords align={vertical},
    ]
\addplot coordinates {(OnlyCloud,1.640) (Mapping,0.100) (Edge-ward,0.100)};
\addplot coordinates {(OnlyCloud,1.685) (Mapping,0.100) (Edge-ward,0.100)};
\addplot coordinates {(OnlyCloud,1.680) (Mapping,0.100) (Edge-ward,1.675)};
\legend{App 1,App 2,App3 }
\end{axis}

\end{tikzpicture}
\caption{Energy Consumption on Cloud.} \label{chart4}
\end{figure}

In this batch of experiments the aim is to understand if, when compared with traditional cloud approaches, the selected data placement algorithms can provide a reduction in energy consumption, thus becoming even more relevant from a perspective of delegation of cloud functions.
Figure \ref{chart4} provides results obtained for energy consumption in Joules for each application, when running each of the three different selected algorithms. As has already been described in related literature \cite{mahmud2018cloud}: the higher the processing in the cloud, the longer the resulting in terms of execution time.

A first observation to be made from the results obtained, is that from the three different algorithms, the one that has lower impact in terms of energy consumption in the cloud is the Mapping algorithm.

A second observation to be made is that storage is the module that affects the most the energy consumption in the cloud. This is relevant in particular for algorithms such as Mapping, which only recur to the cloud for storage.

A third observation is that the energy consumption with Cloud-only increases with an increase in the number of field-level devices. Application 3, which is an example of an intensive application, is accommodated in the Cloud for the cases of the Cloud-only and Edge-ward algorithms. While the Mapping algorithm incurs in a lower cloud energy consumption due to accommodating queuing in Fog devices. 

\section{Conclusions and Next Steps}
\label{conclusions}
This paper describes and evaluates, via simulations, three different data placement algorithms for Fog environments: Cloud-only, Mapping, Edge-ward. The performance evaluation concerns overall execution time, and energy consumption on the Cloud, for low, average, high intensity applications. Three different Fog topologies are also considered.

Results obtained show that the Mapping algorithm consistently achieves the best performance in almost all of the executed simulations, in what concerns execution time. Still, under specific situations, the Edge-ward algorithm can outperform Mapping. This has been observed, for instance, in scenarios where selected Fog devices could not match the processing needs of specific applications, such as occurs with Application 3 (standing for a high intensity application). 

The results of the paper can be used as a micro-benchmark for research related to the execution of IoT applications based on Fog computing. 

As next step, we are developing novel algorithms that take into consideration context-awareness to provide a better selection of edges, based on a concurrent strategy, and taking into consideration opportunistic edge computing frameworks.

\bibliographystyle{IEEEtran}
\bibliography{sample-base}

\end{document}